\documentclass[oneside]{ar-1col_hacked}
\usepackage{url}
\usepackage[numbers]{natbib}
\usepackage{amsmath,amssymb}
\usepackage{hyperref}
\usepackage[framemethod=TikZ]{mdframed}
\usepackage{amsthm}
\usepackage{multirow}
\usepackage{rotating}
\usepackage{soul}
\newcommand{\be}{\begin{equation}}
\newcommand{\ee}{\end{equation}}

\newcommand{\Ap}{A^\prime}
\newcommand{\mAp}{m_{A^\prime}}

\newcommand{\eps}{\epsilon}

\newcommand{\p}{\prime}

\newcommand{\grad}{\nabla}

\newcommand{\order}[1]{\mathcal{O}{(#1)}}
\newcommand{\Eq}[1]{Eq.~\ref{eq:#1}}

\newcommand{\Eqs}[2]{Eqs.~\ref{eq:#1} and \ref{eq:#2}} 
\newcommand{\Sec}[1]{Sec.~\ref{sec:#1}}

\newcommand{\Fig}[1]{Fig.~\ref{fig:#1}}

\newcommand{\bebox}{\begin{empheq}[box=\fcolorbox{light-gray}{light-gray}]{align}}
\newcommand{\eebox}{\end{empheq}}

\newcommand{\GHz}{\text{GHz}}
\newcommand{\THz}{\text{THz}}

\newcommand{\rhodm}{\rho_{_\text{DM}}}
\newcommand{\DM}{{_\text{DM}}}

\newcommand{\dt}{\partial_t}

\newcommand{\eV}{\text{eV}}

\newcommand{\GeV}{\text{GeV}}
\newcommand{\TeV}{\text{TeV}}

\newcommand{\MV}{\text{MV}}

\newcommand{\cm}{\text{cm}}

\newcommand{\Hz}{\text{Hz}}

\newcommand{\gagg}{g_{a\gamma\gamma}}

\newcommand{\w}{\omega}

\newcommand{\g}{\gamma}

\newcommand{\A}{\boldsymbol{A}}

\newcommand{\E}{\boldsymbol{E}}
\newcommand{\B}{\boldsymbol{B}}

\newcommand{\Jv}{\boldsymbol{J}}

\newcommand{\Pv}{\boldsymbol{P}}

\newcommand{\xv}{{\bf x}}

\newcommand{\vv}{{\bf v}}

\newcommand{\pv}{{\bf p}}

\newcommand{\sigmav}{\boldsymbol{\sigma}}
\newcommand{\piv}{\boldsymbol{\pi}}

\newcommand{\Uone}{\text{U} (1)}

\jname{}
\jvol{}
\jyear{}
\doi{}

\begin{document}

\markboth{A. Berlin and Y. Kahn}{New Technologies for Axion and Dark Photon Searches}

\title{New Technologies for Axion and Dark Photon Searches}

\author{Asher Berlin$^{1,2}$, Yonatan Kahn$^{2,3,4}$
\affil{$^1$Theoretical Physics Division, Fermi National Accelerator Laboratory, Batavia, IL 60510, USA; email: aberlin@fnal.gov}
\affil{$^2$Superconducting Quantum Materials and Systems Center (SQMS),
Fermi National Accelerator Laboratory, Batavia, IL 60510, USA}
\affil{$^3$Illinois Center for Advanced Studies of the Universe and Department of Physics,
University of Illinois Urbana-Champaign, Urbana, IL 61801, USA}
\affil{$^4$ Department of Physics, University of Toronto, Toronto, ON M5S 1A7, Canada; email: yf.kahn@utoronto.ca}}

\begin{abstract}
The search for dark matter and physics beyond the Standard Model has grown to encompass a highly interdisciplinary approach. In this review, we survey recent searches for light, weakly-coupled particles -- axions and dark photons -- over the past decade, focusing on new experimental results and the incorporation of technologies and techniques from fields as diverse as quantum science, microwave engineering, precision magnetometry, and condensed matter physics. We also review theoretical progress which has been useful in identifying new experimental directions, and identify the areas of most rapid experimental progress and the technological advances required to continue exploring the parameter space for axions and dark photons. 
\end{abstract}

\maketitle

\tableofcontents

\section{Introduction}  

Near the turn of the 21st century, compelling signatures predicted by weak-scale supersymmetric theories fueled the anticipation of new discoveries at upcoming experiments, such as the Large Hadron Collider and the next generation of underground dark matter direct detection experiments~\cite{Jungman:1995df,Bertone:2016nfn,Cirelli:2024ssz}. As a result, much of the activity in the field of high-energy physics was driven by a small set of common paradigms for what might lie beyond the Standard Model. Today, although it is certainly possible that the ongoing operation of such experiments might soon discover new physics near the electroweak ($\sim \TeV$) scale, it may be the case that the majority of their discovery potential has already been exhausted. 

This state of affairs has led to a considerable relaxation in the community's priors for where new physics might first be unveiled~\cite{Butler:2023glv}. For instance, although it would be theoretically appealing if dark matter were found to be related to other fundamental problems of the Standard Model such as the hierarchy problem, there is no first-principles reason for such a connection. It is also possible that new physics at high energies is well beyond the reach of even the most powerful future colliders. However, even if this were true, dynamics at extremely high energies can give rise to new feebly-coupled low-energy degrees of freedom, motivating observational signatures that are accessible to small-scale \emph{precision} experiments. 
Two examples of such hypothetical particles, and the focus of this review, are the ``axion" and ``dark photon," i.e., dark sector analogs of the ordinary pion and photon that are ubiquitous in theories involving extra dimensions and gauge-coupling unification~\cite{Svrcek:2006yi,Arvanitaki:2009fg,Halverson:2018vbo}.

Driven by these shifts in priors and the thirst for data, many high-energy physicists, theorists and experimentalists alike, have become deeply involved in conceiving and developing small-scale probes for low-energy signatures of new physics~\cite{Hochberg:2022apz,Essig:2022yzw}. These endeavors span a plethora of different subfields involving connections between condensed matter physics, atomic physics, and quantum information science. Compared to two decades ago, the high-energy physics community therefore finds itself in a healthy state of increased diversification.

In this review, we aim to provide a useful entry point for non-experts interested in laboratory precision probes of ultralight axions and dark photons. In the past twenty years, there have been multiple articles of this type (see, e.g., Refs.~\cite{Asztalos:2006kz,Graham:2015ouw}), which have surveyed then-current developments of the most well-known experimental approaches, such as cavity haloscopes, helioscopes, and light-shining-through-wall experiments. Since then, these experiments have improved both their sensitivity and overall scope, but the past decade has also seen an explosion of completely new experimental ideas. Two major developments leveraged by several of these experiments are quantum sensing technology (including squeezed vacuum states and non-demolition measurements such as single-photon counting) and the maturation of high-$Q$ superconducting radiofrequency (SRF) cavities, originally designed for accelerator applications and now harnessed in the search for new physics.

An additional goal of this review is to provide an update on the impressive experimental progress within this diverse field, as well as to identify recent changes in the community's theory priors. Despite this progress, there still remain large gaps in coverage, i.e., regions of theory space that will not be covered by any current proposal. We therefore hope to motivate young theorists and experimentalists to seek the new ideas required to tackle these elusive regions of model space. As we aim to provide a high-level overview, we focus more on experimental concepts than technical details, and refer the reader to the references provided for further study.

The next section begins with a more-detailed overview of the theoretical motivations, priors, and target regions of parameter space for the models considered here. The remaining sections will then discuss laboratory-based searches for ultralight axions and dark photons, depending on whether they do or do not constitute the dark matter of our galaxy. While we unfortunately do not have space in this review to cover the plethora of astrophysical or cosmological probes of axions and dark photons, we refer the reader to the excellent compendia in Refs.~\cite{Baryakhtar:2022hbu,Dvorkin:2022jyg,Boveia:2022syt,Drlica-Wagner:2022lbd}. Throughout, we use a mostly-negative spacetime metric $(+,-,-,-)$ and natural units wherein $\hbar = c = k_B = 1$. 

\section{Theory}
\label{sec:theory}

\subsection{Motivation}  

As mentioned above, axions and dark photons can be motivated by top-down considerations such as extra dimensions and gauge-coupling unification, but they are phenomenologically important because these particles can also explain the dark matter of our universe. In this section, we discuss such bottom-up perspectives, as well as outline the various ways in which axions and dark photons couple to Standard Model fields.

Although the observational evidence for dark matter is overwhelming, direct evidence of its particle properties is almost negligible. However, much can still be learned from basic astrophysical observations and simple theoretical requirements. Taking dark matter to be described by a Poincar\'e-invariant quantum field theory, it is either a boson or fermion. Then, quantum mechanics can be applied on galactic scales to place a lower bound on the dark matter mass $m_\DM$ by demanding that dark matter dominated objects, such as dwarf galaxies, are gravitationally stable. In the case of fermionic dark matter, the Pauli exclusion principle implies that the Fermi velocity does not exceed the gravitational escape velocity of such systems for $m_\DM \gtrsim 100 \ \eV$~\cite{Alvey:2020xsk}. Therefore, sub-eV dark matter must be bosonic. A lower bound on the mass of bosonic dark matter can be derived from the Heisenberg uncertainty principle, which imposes a minimum velocity dispersion for spatially-confined particles. For a galaxy to host a bosonic dark matter halo, the dark matter de Broglie wavelength must be smaller than the size of the galaxy, yielding the constraint $m_\DM \gtrsim 10^{-21} \ \eV - 10^{-19} \ \eV$~\cite{Dalal:2022rmp,Zimmermann:2024xvd}. To proceed further, the interactions of ultralight bosons with $\text{spin} \geq 2$ are severely restricted by Lorentz covariance~\cite{Weinberg:1964ev,Weinberg:1964ew,Weinberg:1965rz,Weinberg:1980kq}. As a result, we are motivated to consider exclusively spin-0 (scalar) or spin-1 (vector) bosons, which are much less constrained and lead to a larger variety of potential signals. 

A massive vector field $A^{\p \, \mu}$ with couplings $\Ap_\mu J^\mu$ to conserved Standard Model currents $J^\mu$ is a well-behaved quantum-mechanical theory; by analogy with the vector field of electromagnetism, we refer to such a field as a dark photon. The situation is not as simple for spin-0 particles. A new spin-0 scalar boson that is invariant under the combination of parity and time-reversal symmetry would have a hierarchy problem even more severe than the Higgs, with quantum threshold corrections from couplings to any heavy particles  tending to push its mass up to the highest possible scale (see, e.g., Ref.~\cite{Craig:2022eqo} for a recent review). On the other hand, a parity-odd pseudoscalar field ``$a$" may be naturally light if it originates as a pseudo-Goldstone boson of a global $\Uone$ symmetry spontaneously-broken at a scale $f_a$, analogous to the pion of QCD. We will refer to such particles as axions throughout the review, reserving the term ``QCD axion'' for a scenario where the spontaneously-broken symmetry is a ``Peccei-Quinn symmetry" responsible for solving the strong-CP problem~\cite{Peccei:1977hh}. This origin for the axion field imposes strong requirements on the structure of its couplings to Standard Model fields, providing an appealing target for direct detection experiments aiming to probe axion interactions with photons, nucleons, and electrons. 

With the motivational underpinnings of axions and dark photons in place, we can now enumerate their interactions with Standard Model fields. Since the axion is a pseudo-Goldstone boson generated by the breaking of a new global $\Uone$ symmetry, its mass and interactions arise analogously to those of the Standard Model pion, such that $m_a \sim \Lambda_a^2 / f_a$ (for the QCD axion, $\Lambda_a$ is roughly the QCD scale $\Lambda_\text{QCD}$). The axion's couplings must respect a continuous shift symmetry, broken to a discrete one only by its small mass and self-interactions. This leads to interactions in the form of higher-dimensional operators suppressed by the scale $f_a$, i.e., $\mathcal{L} \supset J^\mu \, \partial_\mu a / f_a$, where $J^\mu$ is a Standard Model current. Hence, the axion's small mass and coupling-strength are both related to the size of $f_a$, which is constrained to be much larger than the electroweak scale. At energies well below the electroweak scale, these interactions include
\be
\label{eq:axioncouplings}
\mathcal{L} \supset c_\gamma \, \frac{a}{f_a} \, F_{\mu \nu} \, \widetilde{F}^{\mu \nu} + c_g \, \frac{a}{f_a} \, G^a_{\mu \nu} \, \widetilde{G}^{a \, \mu \nu} + c_{\psi} \, \frac{\partial_\mu a}{f_a} \, \bar{\psi} \, \gamma^\mu \gamma^5 \, \psi
~,
\ee
where $F_{\mu \nu}$ and $G_{\mu \nu}^a$ are the photon and gluon field-strengths, respectively; $\widetilde{F}^{\mu \nu}$ and $\widetilde{G}^{a \, \mu \nu}$ are the dual field-strength tensors; $\psi$ is the Dirac field for a Standard Model fermion; and $c_\gamma$, $c_g$, and $c_\psi$ are model-dependent dimensionless coefficients. 

For the spin-1 case, the theory of a vector field coupling to Standard Model currents $J^\mu$ is ill-behaved at high energies unless $J^\mu$ is conserved at the quantum level; otherwise, the coupling of the longitudinal mode is enhanced at high energies, which is strongly constrained by existing searches~\cite{Dror:2017ehi,Dror:2017nsg}. The only fully-conserved (i.e.\ anomaly-free) currents are those corresponding to electromagnetism and baryon-minus-lepton-number. In this work, we focus on the former case, which most naturally arises in models of kinetically-mixed dark photons.\footnote{For a detailed review of dark photon phenomenology across a wide range of masses, see Refs.~\cite{Fabbrichesi:2020wbt,Caputo:2021eaa}.} Here, the $\Ap_\mu$ interacts with the Standard Model photon $A_\mu$ through a renormalizable operator that couples the two field-strengths~\cite{Holdom:1985ag}. In a particular basis choice, this leads to an interaction with the electromagnetic current $J_\text{em}^\mu$ of the form
\be
\label{eq:kinmixdiag}
\mathcal{L} \supset - (A_\mu + \eps \, \Ap_\mu) \, J_\text{em}^\mu 
~,
\ee
where $\eps \ll 1$ is the dimensionless kinetic-mixing parameter. Note that $\eps \ll 1$ is technically natural since $\eps \neq 0$ breaks the combination of dark and visible charge conjugation symmetries to its diagonal subgroup. As we will describe further below, laboratory searches for dark photons are typically easier in their implementation, making the experimental situation quite complementary to those searching for electromagnetically-coupled axions.

\subsection{Targets}  
 
Here, we expand upon the notion of ``theory targets," i.e., regions of parameter space that are theoretically motivated. We begin with the so-called QCD axion. In this case, the axion arises as the Goldstone mode of a new spontaneously-broken Peccei-Quinn global $\text{U}(1)_\text{PQ}$ symmetry~\cite{Peccei:1977ur,Peccei:1977hh,Weinberg:1977ma,Wilczek:1977pj}. New fields coupled to QCD give rise to a mixed-anomaly under  $\text{U}(1)_\text{PQ}$ and the $\text{SU}(3)_\text{C}$ gauge group. In turn, this results in a coupling of the QCD axion to gluons, which compared to \Eq{axioncouplings} we rewrite as 
\be
\label{eq:QCDaxionGG}
\mathcal{L} \supset \frac{a}{f_a} \, \frac{\alpha_s}{8 \pi} \,  G^a_{\mu \nu} \, \widetilde{G}^{a \, \mu \nu}
~,
\ee
where $f_a$ is the reduced axion decay constant and $\alpha_s$ is the strong fine-structure constant. Below the QCD scale, this interaction generates a potential and mass for the axion~\cite{GrillidiCortona:2015jxo},
\be
\label{eq:QCDmass}
m_a \simeq 5.7 \ \mu \eV \times \big( 10^{12} \ \GeV / f_a \big)
~.
\ee
Over cosmological timescales, the axion field relaxes to its minimum at which parity and time-reversal symmetry is preserved. The QCD axion thus provides a dynamical explanation for why the parity and time-reversal violating electric dipole moment (EDM) of the neutron is small, at least 10 orders of magnitude below naive expectations~\cite{Baker:2006ts,Abel:2020pzs}.

Although \Eqs{QCDaxionGG}{QCDmass} imply a direct relationship between the mass $m_a$ and the couplings proportional to $1/f_a$ for the QCD axion, there are models which modify this relationship. For instance, introducing $\mathcal{N} \gg 1$ copies of the Standard Model (and axion) that obey a $\mathbb{Z}_\mathcal{N}$ exchange symmetry gives a QCD-axion with values of $m_a \, f_a$ that are exponentially smaller by $2^{-\mathcal{N}}$, compared to \Eq{QCDmass}~\cite{Hook:2018jle,DiLuzio:2021pxd,DiLuzio:2021gos}. However, this comes at the cost of a $1/\mathcal{N}$ linear tuning, which selects our sector as the one where parity and time-reversal symmetry violation is dynamically relaxed to zero.

Compared to the gluon coupling in \Eq{QCDaxionGG}, the QCD axion's coupling to photons is more model-dependent. This interaction has contributions both from \Eq{QCDaxionGG}, which induces a mixing between the axion and the neutral pion, as well as any additional field content charged under $\text{U}(1)_\text{PQ}$ and electromagnetism. Parameterizing the interaction as
\be
\label{eq:gaggdef}
\mathcal{L} \supset - \frac{g_{a \gamma \gamma}}{4} \, a \, F_{\mu \nu} \, \widetilde{F}^{\mu \nu}
~,
\ee
the coupling coefficient $g_{a \gamma \gamma}$ is related to the additional particle content of the model by
\be
\label{eq:gaggEN}
g_{a \gamma \gamma} \simeq \frac{\alpha_\text{em}}{2 \pi \, f_a} \, \bigg( \frac{E}{N} - 1.92 \bigg) \simeq 2 \times 10^{-16} \ \GeV^{-1} \times \, \bigg( \frac{m_a}{1 \ \mu \eV} \bigg) \,  \, \bigg( \frac{E}{N} - 1.92 \bigg)
~,
\ee
where $E$ and $N$ are the integer coefficients of the $\text{U}(1)_\text{PQ} \times U(1)_\text{EM}^2$ and $\text{U}(1)_\text{PQ} \times \text{SU}(3)_\text{C}^2$ anomalies, respectively, the term involving ``$1.92$" is a recent estimate of the contribution from pion mixing~\cite{GrillidiCortona:2015jxo}, and in the second equality we used the relation of \Eq{QCDmass}. 

Two benchmark models for the QCD axion are the KSVZ axion~\cite{Kim:1979if,Shifman:1979if} and DFSZ axion~\cite{Zhitnitsky:1980tq,Dine:1981rt}, corresponding to $E/N = 0$ and $E/N = 8/3$, respectively, yielding
\be
\label{eq:KSVZDFSZ}
|g_{a \gamma \gamma}| \simeq\bigg( \frac{m_a}{1 \ \mu \eV} \bigg) 
\times
\begin{cases}
3.9 \times 10^{-16} \ \GeV^{-1} & (\text{KSVZ} ~,~ E / N = 0)
\\
1.5 \times 10^{-16} \ \GeV^{-1} & (\text{DFSZ} ~,~ E / N = 8/3)
~.
\end{cases}
\ee
Most experiments choose to focus on these two models when defining the range of interesting parameter space for the QCD axion. There are various reasons for this. The first one is largely historical. The KSVZ and DFSZ theories were introduced many decades ago, only a few years after Peccei and Quinn's first paper on the subject~\cite{Peccei:1977hh}, and so are now considered the classic examples of the QCD axion. Another reason is a slight theoretical bias. DFSZ models are representative of many Grand Unified Theories, although there exist many counterexamples. The final reason is due to practical limitations. As we will discuss below, experiments often need to operate many runs in order to scan over a wide range of possible masses. Due to the time-consuming nature of these endeavors and the demands of funding agencies to outline an attainable and well-defined experimental program, collaborations have used the DFSZ prediction as a natural stopping point when defining their scope. However, it is a simple exercise to construct a model with much smaller $g_{a \gamma \gamma}$. For instance, taking $E/N = 2$ gives
\be
\label{eq:EoverN2}
g_{a \gamma \gamma} \simeq 1.6 \times 10^{-17} \ \GeV^{-1} \times (m_a / \mu \eV) \quad (E/N = 2)
~,
\ee
an order of magnitude below the DFSZ prediction.

By contrast, the target theory space for dark photons is less sharply defined. This is due to the fact that unlike the dimension-five axion couplings, dark photons interact through a marginal dimension-four operator, and are therefore only logarithmically sensitive to the scale of new physics that generates this coupling. More concretely, assuming that $\eps$ vanishes at high energies, loops of particles charged under both the dark photon and electromagnetism radiatively generate a kinetic mixing of size
\be
\eps \simeq \frac{e \, e^\p}{16 \pi^2} \, \sum_i Q_i \, Q^\p_i \, \log{\frac{\mu^2}{M_i^2}}
~,
\ee
where $e$ and $e^\p$ are the electromagnetic and $\Ap$ gauge coupling, the sum runs over all particles of mass $M_i$ charged under both sectors with visible and dark charge $Q_i$ and $Q_i^\p$, respectively, and $\mu$ is the renormalization scale~\cite{Holdom:1985ag,Cheung:2009qd}. Barring cancellations, and assuming that $e^\p \sim e$, this motivates kinetic mixing values of $\eps \sim 10^{-3}$, independent of the scale of new physics. However, it is simple to construct theories where $\eps \ll 10^{-3}$~\cite{Gherghetta:2019coi}. This is the case, for instance, if $e^\p \ll e$, or instead if this interaction arises from mixing between a non-abelian and abelian group, such that it is generated by a higher-dimensional operator suppressed by the mass-scale of new particles charged under both sectors~\cite{Arkani-Hamed:2008kxc}. 

Cosmological considerations can also be used to motivate particular regions of parameter space. However, since early universe production mechanisms are not the focus of this review, we simply refer the reader to the recent work of, e.g.,  Refs.~\cite{Blinov:2019rhb,OHare:2024nmr}. While cosmology can motivate particular regions of parameter space (assuming that such particles make up the entirety of the dark matter), it is also possible that the dark matter density in the Solar System deviates drastically from its average galactic value $\rhodm \simeq 0.4 \ \GeV / \cm^3$~\cite{deSalas:2020hbh}. For instance, axion self-interactions can lead to substructure in the form of large overdensities (``miniclusters") surrounded by significant underdensities (``minivoids")~\cite{Hogan:1988mp,Kolb:1994fi,Kolb:1995bu,Arvanitaki:2019rax}. The rate at which Earth would encounter such a minicluster is likely quite low but depends on the details of the theory and formation mechanism, and is subject to large theoretical uncertainties regarding the survival rate of such objects in the Milky Way. Regardless, the possibility that an $\order{1}$ fraction of the dark matter is bound up in rare dense objects could mean that the effective value of $\rhodm$ for terrestrial experiments would be much smaller than the conventional value of $\rhodm = 0.45 \ \GeV / \cm^3$ adopted within the community of axion experimentalists. It is also plausible that axions or dark photons make up a small subcomponent of the dark matter. Either of these cases highlight the importance of exploring signals weaker than that predicted by the canonical homogeneous dark matter scenario.

\section{Dark Matter in the Lab}
\label{sec:DM}

In this section, we outline various laboratory search strategies for axions and dark photons, assuming that they make up the dark matter of our galaxy. Before delving into the details of the various types of signals, we begin with a brief overview of the theoretical description of  ultralight bosonic dark matter in the laboratory. Although the mass of a dark matter particle is largely unconstrained, its mass density can be inferred from astrophysical observations, with an average value in the solar neighborhood $\rhodm \simeq 0.4 \ \GeV / \cm^3$~\cite{deSalas:2020hbh}. As a result, sub-eV dark matter has a much larger number density $n_\DM \simeq \rhodm/m_\DM$ than traditional weak-scale dark matter, which leads to qualitatively different phenomenology in the laboratory. In particular, considering the typical galactic dark matter velocity $v_\DM \sim 10^{-3}$ and the associated de Broglie wavelength $\lambda_\DM \sim 2\pi/(m_\DM v_\DM)$, the dark matter occupancy per de Broglie volume $N_\text{occ} = n_\DM \, \lambda_\DM^3$ is much larger than unity for $m_\DM \ll 30 \ \eV$. 

Sub-eV dark matter thus has macroscopic phase space occupancy, leading to an interesting interplay of classical and quantum effects. Feebly-coupled ultralight axion or dark photon dark matter fields may be described as a nonrelativistic classical field oscillating at the dark matter mass, 
\be
\label{eq:classicalfields}
a \simeq \frac{\sqrt{2 \rhodm}}{m_a} \, \cos{\big(m_a (t - \vv_\DM \cdot \xv)\big)}
~~,~~
|\A^\p| \simeq \frac{\sqrt{2 \rhodm}}{\mAp} \, \cos{\big(\mAp (t - \vv_\DM \cdot \xv)\big)}
~,
\ee
coherent over a length scale and timescale of
\be
\lambda_\DM \simeq 700 \ \text{m} \times \bigg( \frac{1 \ \mu \eV}{m_\DM}\bigg)
~~,~~
\tau_\DM \simeq 3 \  \text{ms} \times \bigg( \frac{1 \ \mu \eV}{m_\DM}\bigg)
~,
\ee
respectively. Similar results are obtained with a full quantum treatment of the axion field using the formalism of density matrices and open quantum systems~\cite{Derevianko:2016vpm,Hook:2018dlk,Bernal:2024hcc,Cheong:2024ose}; one advantage of this formalism is the ability to precisely specify quantities such as the coherence length $\lambda_\DM$ and time $\tau_\DM$, which can be defined through the field autocorrelation function. 

Experiments aiming to detect axion or dark photon dark matter in the lab will almost invariably measure power spectra of, e.g., laboratory fields, which will end up being proportional to the dark matter two-point function. It is worth noting that it is common in the literature to rewrite the coherence time as $\tau_\DM \sim Q_\DM / m_\DM$ where
\be
Q_\DM \sim v_\DM^{-2} \sim 10^6
\ee
is an effective quality factor of the dark matter field. In this sense, $Q_\DM$ does not have anything to do with dissipation; instead, it describes the dark matter field's spectral width $\sim m_\DM / Q_\DM$ in frequency-space. Sub-eV dark matter also offers the appealing prospect of being able to immediately measure the full 3-dimensional velocity distribution of dark matter in our galaxy following a detection~\cite{Knirck:2018knd,Lisanti:2021vij,Foster:2020fln}, as well as daily modulation from the dark photon polarization~\cite{Caputo:2021eaa}. This is to be contrasted with WIMP experiments, which are rare event searches and thus take much more time to map out the velocity distribution.

\subsection{Electromagnetic Searches}

In this section, we review experimental techniques to probe electromagnetically-coupled axion and dark photon dark matter, defined in \Eqs{gaggdef}{kinmixdiag}, respectively. As we will see below, both can lead to effective source terms in Maxwell's equations. Thus, searches for either model are highly synergistic, with the main difference often being that axion experiments typically require applied magnetic fields, whereas dark photon searches do not.

Let us begin with a description of axion electrodynamics. Experimentally, the most readily accessible axion coupling in laboratory experiments is its interaction with photons. Indeed, the earliest experimental searches for axions with sensitivity to $f_a$ well above the electroweak scale were electromagnetic cavity experiments dating back to the late 1980s~\cite{DePanfilis:1987dk,Wuensch:1989sa,Hagmann:1990tj}. In the last decade, there has been a flourishing of new experimental ideas, many of which have already come to fruition and begun taking data.

The effect of the axion-photon coupling $\gagg$ is most easily seen by considering how it changes Maxwell's equations. In  Gauss's and Amp\`{e}re's laws, the axion field enters as an effective charge and current density in the presence of electromagnetic fields,
\be
\label{eq:jeffaxion}
\rho_{a \g} \equiv -\gagg \, \B \cdot \grad a
~~,~~
\Jv_{a \g} \equiv -\gagg \, \big(\E \times \grad a - \B \,  \dt a\big)
~.
\ee
In an experiment searching for axion dark matter, the second term in $\Jv_{a \g}$ dominates, since from \Eq{classicalfields} the spatial gradient of the axion field is suppressed by the small dark matter velocity $v_{\rm DM} \sim 10^{-3}$. Thus, in the presence of a large background laboratory magnetic field $\B_0$, axion dark matter sources an effective current density $\Jv_{a \g} \simeq \gagg \, \B_0 \, \dt a$, which in turn generates small response fields proportional to $\gagg$ across a narrow frequency band of width $m_a / Q_\DM$ near $m_a$. Such signal fields can be significantly enhanced in amplitude through the use of electromagnetic resonant detectors, such as cavities and LC circuits. 

Kinetically-mixed dark photons also modify Maxwell's equations. To describe physical effects, it is often more convenient to change basis from that of \Eq{kinmixdiag} to the so-called ``visible" and ``invisible" field basis, which can be identified by noting that the visible linear combination of fields $A_\text{vis} \simeq A + \eps \, \Ap$ couples to Standard Model currents, whereas the invisible linear combination $A_\text{inv} \simeq \Ap - \eps \, A$ is completely decoupled ($A_\text{vis}$ is the field that is, e.g., screened by electromagnetic shields). The dark photon field sources an effective charge and current density for visible fields~\cite{Graham:2014sha,Berlin:2023mti}
\be
\label{eq:jeffAprime}
\rho_{\Ap} \equiv - \frac{\eps \,  \mAp^2}{1 - \eps^2} \, \, \phi^\p
~~,~~
\Jv_{\Ap} \equiv - \frac{\eps \,  \mAp^2}{1 - \eps^2} \, \, \A^\p
~,
\ee
where we decomposed the dark photon field as $A^{\p \, \mu} = (\phi^\p , \A^\p)$. For a massive vector field, charge continuity demands that $\partial_\mu A^{\p \, \mu} = 0$,\footnote{Note that for a massive vector field, this is actually not a gauge choice, but is instead a consequence of conservation of dark charge $J^{\p \, \mu}$ and arises from the equations of motion for $A'$.}
such that the dark scalar potential is parametrically smaller than the dark vector potential, $\phi^\p \sim v_\DM \, |\A^\p| \ll |\A^\p|$. From the expressions above, we see that the $\Ap$ decouples from the Standard Model in the massless limit $\mAp \to 0$, which is due to the fact that our visible photon is simply the linear combination $A + \eps \Ap$ in \Eq{kinmixdiag} (an exception to this is if there are dark sector particles directly charged under the $\Ap$, as in models of millicharged particles, which couple to both sectors~\cite{Berlin:2022hmt,Berlin:2023gvx}). Hence, any calculation that implies sensitivity to a massless dark photon (in the absence of dark charges) should be met with skepticism.

From \Eq{jeffAprime} we see that the dominant effect for dark photon dark matter is also an effective current density $\Jv_{\Ap} \simeq - \eps \, \mAp^2 \, \A^\p$, now aligned along the polarization of the dark photon field rather than an external magnetic field. Indeed, unlike in the case of the axion, the dark photon signal will appear whether or not the magnetic field is on. As a result, the sensitivity of experiments searching for electromagnetically-coupled axion dark matter can often be recast in terms of dark photon dark matter~\cite{Caputo:2021eaa} (so long as care is taken in regards to any magnetic field veto). Comparing the form of the effective currents in \Eqs{jeffaxion}{jeffAprime} implies that an axion experiment employing a magnetic field of strength $B_0$ that is sensitive to a coupling $\gagg$ is typically sensitive to dark photons of the same mass with a kinetic mixing of $\eps \sim \gagg \, B_0 / m_\DM$.\footnote{A notable exception is a toroidal axion experiment, the geometry of which is maximally mismatched to a dark photon search~\cite{Caputo:2021eaa}.} Especially in cases where the experimental apparatus involves superconducting elements, a dedicated dark photon search without the external $B$-field often involves fewer technical complications.

\begin{figure}[t!]
\includegraphics[width=0.85\textwidth]{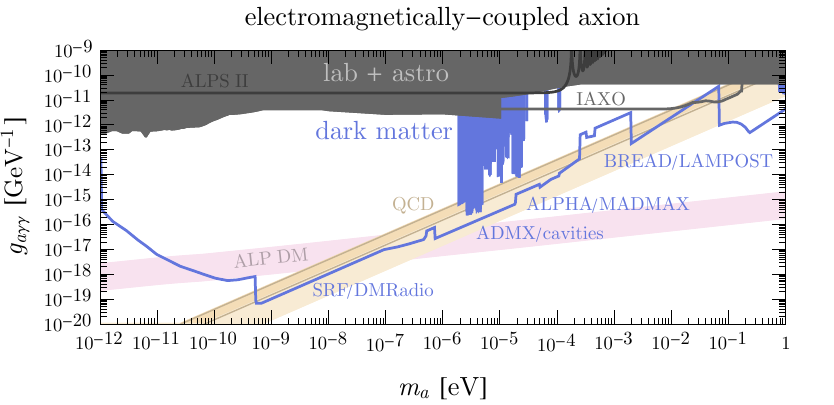}
\caption{A summary of the parameter space for electromagnetically-coupled axions (\Eq{gaggdef}), in the plane spanned by the axion-photon coupling $\gagg$ and axion mass $m_a$. Gray regions or lines do not assume that axions make up the dark matter, whereas blue ones assume axions comprise the full galactic dark matter density. Shaded  regions correspond to existing limits, whereas lines correspond to the projected sensitivity of proposed experiments. Labels refer to experiments discussed throughout this review. The darker orange band denotes the range of couplings and masses as motivated by the canonical QCD axion models of \Eq{KSVZDFSZ}, whereas the bottom of the light orange region corresponds to the QCD axion model of \Eq{EoverN2}. Along
the pink band, axion-like particle production through the misalignment mechanism with an $\order{1}$ initial misalignment angle is consistent with the observed dark matter energy density~\cite{Blinov:2019rhb,OHare:2024nmr}.
\label{fig:axionphoton}}
\end{figure}

The important scale in electromagnetic axion and dark photon dark matter experiments is the dimensionless quantity $m_\DM \, L_\text{exp}$, involving the dark matter mass and the size of the experiment $L_\text{exp}$, which determines the relative importance of retarded-time and radiation effects.\footnote{To be more precise, retarded-time effects are typically governed by the physical distance a current must travel around a device, so for a cylindrical geometry, $L_\text{exp}$ is the circumference rather than the radius (see Ref.~\cite{Benabou:2022qpv} for an example).} Although not every experimental strategy is perfectly captured by the following categorization, three regimes motivate qualitatively different experimental techniques: 1) In the quasi-static regime ($ m_\DM \, L_\text{exp} \ll 1$), retarded-time effects are negligible, such that the dark matter effective current sources a quasi-static signal magnetic field. 2) In the cavity regime ($m_\DM \, L_\text{exp} \sim 1$), retarded-time effects are maximal, i.e., $\dt \E$ and $\grad \times \B$ are comparable in magnitude in Amp\`{e}re's Law. 3) In the radiation regime ($m_\DM \, L_\text{exp} \gg 1$), the displacement current term $\dt \E$ dominates over $\grad \times \B$. 

\begin{figure}[t!]
\includegraphics[width=0.85\textwidth]{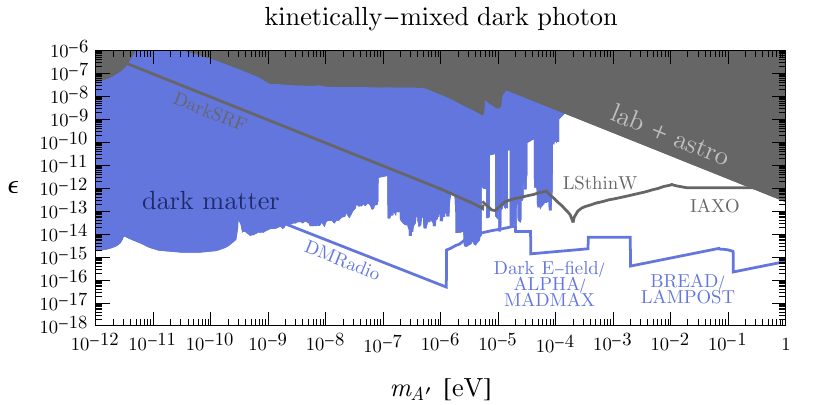}
\caption{As in \Fig{axionphoton}, but for the kinetically-mixed dark photon parameter space (\Eq{kinmixdiag}), spanned by the kinetic mixing parameter $\eps$ and the dark photon mass $\mAp$. 
\label{fig:darkphoton}}
\end{figure}

At a practical level, the boundary between these regimes is set by the size of typical state-of-the-art conducting cavities ($L_\text{exp} \sim \text{cm} - \text{m}$, corresponding to $L_\text{exp}^{-1} \sim \GHz$ frequencies), as well as by the demands of the readout, with GHz corresponding to the transition between DC SQUID readout at lower frequencies and microwave electronics at higher frequencies. As such, we will organize our discussion of dark matter experiments searching for electromagnetic couplings with respect to the GHz scale. A high-level overview of existing limits and projected sensitivities of future experiments is shown in Figs.~\ref{fig:axionphoton} and \ref{fig:darkphoton}.

Finally, let us provide a general parametric estimate for the signal power which can be applied to many of the setups described below. The dark matter effective current $\Jv_\DM = \Jv_{a \g}$ or $\Jv_{\Ap}$ will deposit power into an experiment operating at a frequency $\w_\text{exp}$ (in general, $\w_\text{exp} \neq m_a$). In any of the three frequency regimes, the signal power in many experimental setups can be written schematically as
\be
\label{eq:GenPsig}
P_\text{sig} \sim \big( \eta \, J_\DM \big)^2 \, L_\text{exp}^4 \, \min\big( 1 \, , \, \w_\text{exp} \, L_\text{exp} \big) ~ \min\bigg( Q_\text{exp} \, , \, \frac{\w_\text{exp}}{m_\DM} \, Q_\DM \bigg)
~.
\ee
Here, $\eta \lesssim 1$ is a dimensionless form factor quantifying the overlap between the dark matter field and the experimental setup, such that $\eta \sim \order{1}$ in optimized geometries; note that $\eta$ can be much smaller for, e.g., large open dish antennas  with $\eta \sim (\w_\text{exp} \, L_\text{exp})^{-1} \ll 1$. $Q_\text{exp}$ is the quality factor of the detector, which is $\order{1}$ for open setups, but can be $\gg 1$ for cavity or lumped-element resonators. Furthermore, we have ignored details pertaining to the aspect ratio of the system; for instance, for experiments consisting of spaced layers of dielectrics, $L_\text{exp}^4$ should be replaced by $\text{area} \times \text{length}^2$ in \Eq{GenPsig}.  Although the integrated signal power in \Eq{GenPsig} is independent of the detector's quality factor for $Q_\text{exp} \gtrsim (\w_\text{exp} / m_\DM) \, Q_\DM$, the sensitivity of a resonant setup is often still enhanced for even larger $Q_\text{exp}$. This is due to the fact that the total noise power integrated over the resonator bandwidth $\sim \w_\text{exp} / Q_\text{exp}$ is suppressed compared to $P_\text{sig}$ if, e.g., the noise temperature is independent of $Q_\text{exp}$. In this case, the signal-to-noise ratio generally scales as the geometric mean $\sqrt{Q_\text{exp} \, Q_\DM}\, $ for a fixed scanning time to cover an $e$-fold in dark matter mass. 

%

\subsubsection{GHz frequencies}
\label{sec:axionphotonGHz}

The first experiments that had plausible direct detection sensitivity to dark matter axions were GHz microwave cavity experiments dubbed ``haloscopes''~\cite{DePanfilis:1987dk,Wuensch:1989sa,Hagmann:1990tj}, which have evolved into the modern ADMX program and influenced the design of other newer experiments. These experiments employ magnetic fields applied to cavities tuned to the dark matter frequency, such that $\w_\text{exp} \sim L_\text{exp}^{-1} \sim m_\DM$. Independent of the magnetic field, these setups have sensitivity to dark photon dark matter as well. The narrow-band nature of such experiments requires constant operations to tune to new frequencies, and often the experimental focus is on maintaining live-time while collecting data from a system that has many moving components. The key figure of merit for cavity experiments is the dark matter frequency scan rate,
\be
\label{eq:scanrate}
\frac{d\w_\text{exp}}{dt} \propto Q_L \, Q_\DM (\rhodm \, V_\text{exp} \, \eta / T_n)^2 \times 
\begin{cases}  (\gagg \, B_0)^4  & \text{(axion)}
\\
\eps^4 & \text{(dark photon)}
~,
\end{cases}
\ee
where $Q_L$ is the loaded cavity quality factor, $V_\text{exp}$ is the cavity volume, $\eta$ is a mode-overlap form factor, and $T_n$ is the effective noise temperature. It is important to note that \Eq{scanrate} holds even when $Q_L \gg Q_\DM$; consistent with \Eq{GenPsig}, there is no penalty for having an intrinsic cavity linewidth narrower than the dark matter, so long as tuning steps are taken with fractional width $Q_\DM^{-1}$ rather than $Q_L^{-1}$~\cite{Chaudhuri:2018rqn,Berlin:2019ahk,Cervantes:2022gtv}. 

Over the last decade there has been a steady progression of experiments that have come online, and new ones are progressing further with R\&D demonstrator and conceptual designs. Two experiments, ADMX and CAPP, have matured to the point of searching down to the QCD line over a fairly wide frequency range. As of this writing, ADMX-G2 has covered $(0.65-1.02)\ \GHz$~\cite{PhysRevLett.120.151301,PhysRevLett.124.101303,PhysRevLett.127.261803} and CAPP has covered $(1.025-1.185) \ \GHz$~\cite{PhysRevX.14.031023} at or near DFSZ sensitivity. Both experiments plan to continue scanning with existing technology: copper cavities with one or more tuning rods, large-bore magnets with modest fields (NbTi at $7.8  \ \text{T}$ for ADMX, and $10.8 \ \text{T}$ for CAPP-MAX), and cryogenic amplification with current-pumped or flux-pumped Josephson Parametric Amplifiers (JPAs) running as a linear amplifier near the quantum limit.

One of the primary research directions that has borne fruit is the development of field-tolerant SRF cavities. Though superconducting accelerator cavities have quality factors as large as $Q \gtrsim 10^{11}$~\cite{Romanenko:2014yaa}, it is generally believed that RF losses due to flux vortex motion degrade surface loses at high-field to below that of copper. Fortunately, these concerns are not relevant for dark photon searches which do not require an external $B$-field, and Ref.~\cite{Cervantes:2022gtv} recently set the strongest constraints on $\eps$ at any dark photon mass using a cavity with $Q_L = 4.7 \times 10^9$, obtaining $\eps < 1.5 \times 10^{-16}$ in a narrow band around $\mAp = 5.35 \ \mu \eV$. That said, SRF cavities made from Nb$_3$Sn films demonstrated $Q \sim 5 \times 10^5$ in a $6 \ \text{T}$ magnetic field~\cite{Posen:2022tbs}, and work is continuing to use Nb$_3$Sn coatings to increase $Q$ for superconducting cavities in high fields. Other ideas include a variety of material coatings on various substrates (for example, NbTi on Cu, Nb3Sn on Nb, Nb3Sn on Cu, or ReBCO high-temperature superconductor tape), which have been demonstrated in empty cavities~\cite{Alesini:2019ajt,Belomestnykh:2022scw,Yoon:2022gzp}. Tunable systems using these coatings are beginning to be deployed~\cite{Ahn:2019nfy,Ahn:2019gbo} with quality factors approaching $\sim 10^6$ being demonstrated.

\subsubsection{Low Frequencies (below GHz)}
\label{sec:kHzGHz}

Since the lowest resonant frequency of a cavity scales as its inverse size, probing sub-$\mu \eV$ masses with the cavity haloscopes discussed above requires prohibitively large systems. Various experimental schemes have been proposed to overcome this. 

One strategy involves the use of lumped-element LC circuits as electromagnetic resonators; their resonant frequency is not directly tied to their inverse geometric size, allowing them to operate in the quasi-static limit, $\w_\text{exp} = 1 / \sqrt{L C} \ll 1/L_\text{exp}$. As for cavity haloscopes, such a setup requires tuning the resonant frequency to the dark matter mass $\w_\text{exp} \simeq m_\DM$, as well as employing an external magnetic field to search for axion dark matter. A large pickup inductor can be used to resonantly amplify the magnetic field sourced by either the axion or dark photon effective current. This approach is being undertaken by the DMRadio collaboration~\cite{DMRadio:2022pkf}, which plans on developing large LC resonators with quality factors of $Q_\text{LC} \sim Q_\DM \sim 10^6$ with the ultimate goal of probing QCD axion dark matter in the  $\text{MHz} - \text{GHz}$ frequency range. Implementing this strategy requires a large high-$Q$ sapphire capacitor which is tunable over decades of capacitance to parts-per-million precision, which DMRadio is currently developing. The ADMX SLIC experiment has demonstrated the viability of this strategy over a narrow frequency range, using piezoelectric tuning of an LC circuit inside a solenoidal magnet~\cite{Crisosto:2019fcj}. At the highest frequencies, approaching the cavity regime with $m_\DM \, L_\text{exp} \sim 1$, the lumped-element approximation breaks down and inductors develop a capacitive impedance~\cite{Benabou:2022qpv}, which induces parasitic resonances and requires several inductive pickup sheaths of various sizes to fully cover the desired mass range~\cite{DMRadio:2023igr}. In the absence of a tunable capacitor, one can still use purely inductive readouts to perform a broadband search, as was pioneered by ABRACADABRA~\cite{Salemi:2021gck} and SHAFT~\cite{Gramolin:2020ict} (a non-tunable capacitor would improve sensitivity even off-resonance~\cite{Chaudhuri:2018rqn}).

Heterodyne searches based on frequency up-conversion schemes have also been proposed in Refs.~\cite{Berlin:2019ahk, Lasenby:2019prg, Berlin:2020vrk} to probe sub-GHz QCD axion dark matter. These setups employ a SRF cavity with two nearly-degenerate tunable modes, referred to as the “pump mode” and “signal mode.” The cavity pump mode is driven at angular frequency $\w_0 \sim \text{GHz}$. Since the axion effective current $\Jv_{a \g}$ involves the product of the pump magnetic field and the axion field, $\Jv_{a \g}$ oscillates at the beat frequencies $\w_\text{exp} = \w_0 \pm m_a$. Thus, if the signal mode is tuned to the nearby frequency $\w_1 \simeq \w_0 \pm \Delta \w$ offset by $\Delta \w \ll \w_0$, then axion dark matter oscillating at the frequency splitting  $m_a \simeq \Delta \w$ can resonantly drive power from the pump to the signal mode. This approach is currently being pursued at Fermilab~\cite{Giaccone:2022pke}. Compared to static-field lumped-element resonators, the main qualitative difference here arises from the fact that the applied magnetic field is in the form of a time-dependent driven mode of the cavity, which upconverts the low-frequency axion into a higher frequency RF signal, $\w_\text{exp} \gg m_a$. As can be inferred from \Eq{GenPsig}, this has two important implications. First, in general the integrated signal power $P_\text{sig}$ saturates once the coherence time of the resonator overcomes that of the axion field. In an up-conversion setup, this occurs  once $Q_\text{exp} \gtrsim (\GHz / m_a) \, Q_a \gg 10^6$, such that $P_\text{sig}$ is enhanced by the large quality factors achievable with SRF cavities (as large as $Q_\text{exp} \sim \text{few} \times 10^{11}$)~\cite{Romanenko:2014yaa}. Second,  Lenz's law implies that $P_\text{sig}$ is dictated by $\dt J_{a \g}$. Thus, the signal power in a static-field setup is suppressed by the small axion mass $P_\text{sig} \propto m_a$ for $m_a \ll L_\text{exp}^{-1} \sim \mu \text{eV}$. However, by oscillating the applied field in an up-conversion experiment, this suppression is undone. 

Either of these low-frequency search strategies needs to contend with an irreducible background arising from the large number of thermal photons at finite temperature. In this case, thermal fluctuations preclude the usefulness of techniques such as photon counting and squeezing. However, for a scanning experiment, lowering the readout noise well below the standard quantum limit (which dominates off-resonance) enhances the signal-to-noise ratio by broadening the so-called ``sensitivity bandwidth" to be much greater than the resonator bandwidth~\cite{Chaudhuri:2018rqn,Berlin:2019ahk,Chaudhuri:2019ntz}. Thus, as opposed to photon counting, quantum metrology related to low-noise phase-sensitive readout of voltage~\cite{Kuenstner:2022gyc}, which allows the frequency to be determined, could potentially further enhance the scan rate of these setups. 

An effect closely related to the heterodyne one arises from modifications to the photon's dispersion relation. In an axion dark matter background, the phase velocity of left- and right-polarized light with frequency $\w_0$ is shifted by $\delta v \simeq \pm \gagg \, \sqrt{2 \rhodm} / (2 \w_0)$, which is equivalent to a rotation of linear polarized light. Experimental setups based on optical interferometry have been proposed in, e.g., Refs.~\cite{DeRocco:2018jwe,Liu:2018icu}, which rely on the phase difference accumulated by the different circular polarizations. Such schemes could cover new parameter space 1--2 orders of magnitude beyond existing astrophysical bounds.

\subsubsection{High frequencies (GHz and above)}
\label{sec:HF}

At high frequencies, at least two challenges for the resonant cavity approach become apparent. First, from the scan rate in \Eq{scanrate}, the form factor $\eta$ is $\order{1}$ only when the dark matter Compton wavelength is of order the cavity size, which requires a cavity volume of $V_\text{exp} \sim 1/m_\DM^3$, limiting the sensitivity for larger masses. Second, the parametric amplifiers used for cavity searches no longer operate at THz frequencies and above, so a qualitatively different readout is required. Here, we discuss approaches to mitigate these challenges.

At frequencies not too far above 1 GHz, the loss in sensitivity from the reduced volume of a single smaller cavity can be counteracted with lower readout noise. Strategies based on quantum metrology are actively being developed in this frequency range, which may also prove more generally useful in other  regimes. For instance, HAYSTAC~\cite{HAYSTAC:2020kwv,HAYSTAC:2023cam} has pioneered the use of low-noise squeezed-state JPAs to increase the effective cavity bandwidth, thus enhancing the scan rate at each frequency step beyond that implied by the standard quantum limit. Currently, HAYSTAC is the only axion experiment to have demonstrated operation beyond the standard quantum limit, and the factor of $\sim 2-3$  in scan rate gained by squeezing is currently limited by the state of quantum technology, offering a promising opportunity for improvement in the future. HAYSTAC has already taken and analyzed data in the $(4-5) \ \GHz$ frequency range, and expects to cover $(10-12) \ \GHz$ in future runs.  Other approaches involve single-photon detection~\cite{Dixit:2020ymh} as a readout strategy, where the occupation number of photons in a cavity may be non-destructively measured through its effect on the frequency of a qubit oscillation. Indeed, Ref.~\cite{Dixit:2020ymh} has already demonstrated single-photon detection as a viable strategy for dark photon detection, constraining $\eps < 1.7 \times 10^{-15}$ for dark photon dark matter in a narrow band around $\mAp \simeq 24.86 \ \mu \eV$. An alternate strategy involves quantum-mechanically enhancing the signal by preparing a superconducting qubit in a non-classical Fock state~\cite{Agrawal:2023umy}. The challenge of extending these latter strategies to axion detection involves maintaining qubit operation near a large magnetic field~\cite{Bal:2023ccn}.
 
Another approach involves maintaining a large effective volume at higher frequencies. For instance, ADMX-EFR~\cite{ADMXEFR} proposes to read out 18 small cavities simultaneously in a $9.4 \ \text{T}$ magnet to cover the $(2-4) \ \GHz$ frequency range. One may also consider creating a ``metamaterial'' consisting of a dense array of conducting wires in order to modify the photon plasma frequency $\omega_p$, thus facilitating resonant conversion when $\omega_p \sim m_\DM$ and decoupling the experimental volume from the form factor~\cite{Lawson:2019brd}. The ALPHA experiment will use this ``plasma haloscope'' setup to search in the $(5-50) \ \GHz$ range; performance of the metamaterial has been validated recently at $\sim 10$ GHz~\cite{Wooten:2022vpj}.

Moving to higher frequencies, MADMAX~\cite{MADMAX:2019pub} and DALI~\cite{DeMiguel:2023nmz} use an array of dielectric disks to exploit electromagnetic boundary conditions. In the presence of the dark matter effective current, electric fields must be induced at the disk boundary, producing radiation at a frequency $\omega_\text{exp} \simeq m_\DM$. Many such disks may be placed inside a coherence length, allowing the radiation to constructively interfere and generating either a broadband or narrow-band response function depending on the disc spacing. For instance, operating in the so-called ``transparent mode," $N$ disks of refractive index $n$ and thickness $\pi / (n \, m_\DM)$ are separated by a vacuum gap of thickness $\pi / m_\DM$. In this case, the outgoing radiation adds coherently, and the sensitivity bandwidth of the setup is roughly $\sim m_\DM / N$. Demanding that the total length of the experiment fits inside a coherence length ($N \lesssim 1 / v_\DM \sim 10^3$) then leads to the optimal signal power scaling in \Eq{GenPsig}.  A MADMAX prototype with three sapphire disks in a $1.6 \ \text{T}$ $B$-field has recently taken first data, demonstrating a power boost factor of $\sim 2500$ from constructive interference from the disks and constraining an axion coupling of $\gagg < 2 \times 10^{-11} \ \GeV^{-1}$ in a narrow range of masses around $m_a \simeq 80 \ \mu \eV$~\cite{MADMAX:2024jnp}. This prototype was operated in the 10 GHz range and was thus able to use a similar readout to HAYSTAC; however, the full experiment, which aims for $(10 - 100) \ \GHz$, will likely need new readout strategies beyond the quantum limit at the highest frequencies~\cite{MADMAX:2019pub}. 

A similar strategy can be employed with much smaller dielectric spacing, read out with superconducting nanowire single-photon detectors (SNSPDs) to probe near-optical frequencies. This is the strategy pursued by, e.g., LAMPOST~\cite{Baryakhtar:2018doz}, which leverages SNSPDs with extremely low dark rates in a commensurately scaled-down geometry involving several $\mu{\rm m}$-thick alternating layers fused into a ``dielectric stack." First data has been taken without the magnetic field, constraining $\eps \lesssim 10^{-12}$ for dark photon masses of $\mAp \sim (0.7 - 0.8) \ \eV$~\cite{Chiles:2021gxk}; the dark rates for the SNSPDs were an impressive $6 \times 10^{-6} \ \Hz$, corresponding to 4 total counts over the 180-hour exposure. Future runs aim to push toward masses of $10 \ \text{meV}$, which is within the sensitivity band of lower-threshold SNSPDs but with potentially higher dark rates. A similar approach, MuDHI, uses a 23-layer dielectric stack with a single-photon avalanche detector at $1.5 \ \eV$ and has set dark photon limits at $\eps \lesssim 10^{-10}$~\cite{Manenti:2021whp}. 

Other approaches have eschewed resonant scanning in favor of fundamentally broadband setups with sensitive photon detectors. For instance, the BREAD collaboration has revived an older approach using a dish antenna in a specialized parabolic geometry that focuses the dark matter-induced radiation onto a small detector area~\cite{BREAD:2021tpx}. Multiple readout techniques, including kinetic inductance detectors at $(0.1 - 1) \ \THz$, quantum capacitance detectors at $(1-10) \ \THz$, and SNSPDs at $(10-100) \ \THz$, are envisioned to cover several decades of axion masses down to the KSVZ target and potentially beyond. Broadband approaches have also been developed at $\GHz - \THz$ frequencies, such as the  Dark $E$-field pilot experiment~\cite{Godfrey:2021tvs}, which used a dipole antenna in a shielded room  to search for the electric field generated by dark photon dark matter.

Another broadband approach at higher frequencies leverages the development of low-noise targets for the detection of sub-GeV dark matter scattering. In particular, target materials hosting in-medium electronic~\cite{Berlin:2023ppd} and phonon~\cite{Mitridate:2020kly} excitations, are also sensitive to axion dark matter if these detectors are operated in strong magnetic fields, thus creating a ``magnetized medium." In this case, the total signal rate is governed by the dielectric energy-loss function $P_\text{sig} \propto \text{Im} (- \varepsilon^{-1})$ involving the permittivity evaluated at frequency $\w \simeq m_a$~\cite{Berlin:2023ppd}, demonstrating that sensitivity to new parameter space is possible for $m_a \sim (0.1 - 10) \ \eV$, provided that existing dark counts are reduced compared to existing values.

\subsection{QCD}
\label{sec:QCD}

\begin{figure}[t]
\includegraphics[width=0.85\textwidth]{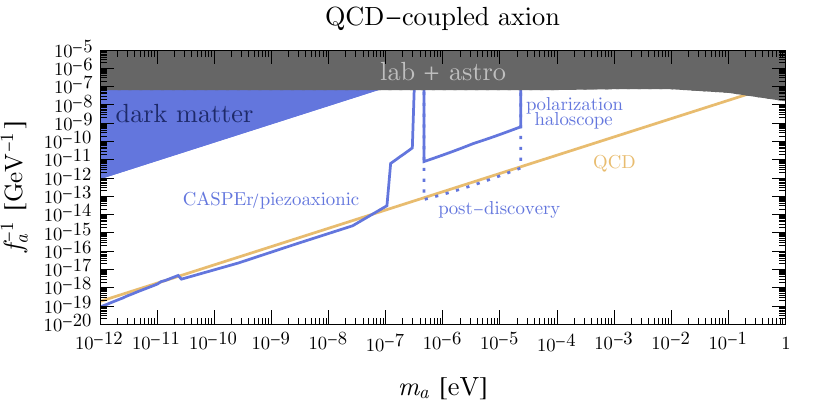}
\caption{As in \Fig{axionphoton}, but for the defining coupling of the QCD axion (\Eq{QCDaxionGG}), in the parameter space spanned by the inverse axion decay constant $f_a^{-1}$ and the axion mass $m_a$. The orange line denotes the standard QCD axion relation of \Eq{QCDmass}. The dotted blue line shows  the reach for a polarization haloscope experiment which targets a single candidate QCD axion mass~\cite{Berlin:2022mia}; this would be the case to definitively test whether a signal at an RF haloscope does indeed arise from the QCD axion.
\label{fig:QCDaxion}}
\end{figure}

Here, we give an overview of existing and proposed dark matter searches for the defining coupling of the QCD axion to gluons, \Eq{QCDaxionGG}. A summary of existing limits and projected sensitivities is shown in \Fig{QCDaxion}. For this coupling, the dark matter axion field  is equivalent to an effective QCD theta angle oscillating at a frequency set by the dark matter mass,
\be
\label{eq:thetaQCD}
\theta_a = \frac{a}{f_a} \simeq \frac{\sqrt{2 \rhodm}}{m_a \, f_a} \, \cos{m_a t} \simeq 4.3 \times 10^{-19} \, \cos{m_a t}
~,
\ee
where we used \Eq{QCDmass} in the second equality. This gives rise to various effects that violate parity and time-reversal symmetry at low energies. One such effect is that the neutron acquires a non-vanishing \emph{oscillating} electric dipole moment, $d_n \simeq \big( 2.4 \times 10^{-3} \ e \ \text{fm} \big) \, \theta_a$, which generates a coupling between the neutron spin and an applied electric field $\E_0$.

The CASPEr-Electric experiment~\cite{Budker:2013hfa,JacksonKimball:2017elr} aims to search for this interaction with nuclear magnetic resonance (NMR) techniques. The induced electric dipole moment leads to a small oscillating nuclear polarization aligned with a background electric field $\E_0$, which can be read out with a precision electromagnetic sensor, such as a SQUID. For nuclear spins initially polarized along an applied magnetic field $\B_0$ transverse to $\E_0$, the spin-response can be resonantly enhanced if the dark matter frequency matches the Larmor frequency, $m_a \simeq \g_n B_0$, where $\g_n$ is the gyromagnetic ratio of the nucleon. As a result, this setup can scan over a range of axion masses by tuning $B_0$. The overall effect can be further amplified with the use of ferroelectric materials with large internal electric fields. The ultimate spin projection noise-limited reach of CASPEr-Electric can explore QCD axion dark matter in the $(10^{-12} - 10^{-7}) \ \eV$ mass range (see also Ref.~\cite{Dror:2022xpi}). A similar range of QCD axion masses could also potentially be covered by an experiment designed to search for the so-called ``piezoaxionic" effect~\cite{Arvanitaki:2021wjk}. In this case, the oscillating nuclear electric dipole moment couples to the mechanical stress of a nuclear spin-polarized piezoelectric crystal, which can be resonantly enhanced when the axion mass is close to a vibrational mode of the system. 

Both of the aforementioned strategies are optimized for sub-GHz frequencies, corresponding to axion masses $m_a \lesssim 10^{-6} \ \eV$. In the case of CASPEr-Electric, this is due to the fact that resonance is achieved for $m_a \simeq \g_n \, B_0$, such that probing $m_a \gtrsim 10^{-6} \ \eV$ would require magnetic fields greater than those accessible in the laboratory, $B_0 \gtrsim 10 \ \text{T}$. Furthermore, an experiment based on the piezoaxionic effect is optimal when $m_a$ is comparable to a low-lying mechanical resonance, which for a large sample typically lies well below  $\order{1} \ \GHz$ in frequency, due to the small speed of sound. To probe $\gtrsim \GHz$ frequencies, the so-called ``polarization haloscope" has been proposed~\cite{Berlin:2022mia}, which involves placing a nuclear spin-polarized dielectric in a conducting cavity. Analogous to the neutron electric dipole moment, \Eq{thetaQCD} generates an oscillating atomic electric dipole moment $d_A$ directed along the spin. The corresponding oscillating polarization density $P_\text{EDM} \sim n_A \, d_A$ in a medium of atomic spin density $n_A$ induces an electromagnetic current $\Jv_\text{EDM} = \dt \Pv_\text{EDM}$, which can resonantly excite the mode of a microwave cavity whose frequency is comparable to the axion mass. For the QCD axion, comparison between $\Jv_\text{EDM}$ and $\Jv_{a \g}$ implies that the signal generated by the QCD coupling is typically a few orders of magnitude smaller~\cite{Berlin:2022mia}. Regardless, a polarization haloscope could explore new parameter space at high frequencies, as well as definitively test whether a potential future signal seen at an experiment such as ADMX indeed arises from the QCD axion. In any of the experiments discussed in this section, achieving the sensitivity shown requires overcoming various experimental difficulties. This includes minimizing noise at low frequencies, as well as identifying optimal materials that can be prepared with large nuclear spin-polarization fractions and enhanced Schiff moments (thus allowing large internal electric fields).

Before concluding this section, we also note that the QCD axion can lead to effects that obey parity and time-reversal symmetry, but at higher-order in $\theta_a$. For example, at $\order{\theta_a^2}$, there exist corrections to nucleon and pion masses~\cite{Blum:2014vsa}, as well as the parity and time-reversal even interaction with photons, $\mathcal{L} \supset \order{10^{-4}} \, \theta_a^2 \, F_{\mu \nu} F^{\mu \nu}$~\cite{Kim:2023pvt,Beadle:2023flm}. This interaction modifies the effective fine-structure constant, thus shifting atomic energy levels and potentially generating signals that can be searched for with atomic clock experiments. Since this interaction is parametrically suppressed compared to those linear in $\theta_a$, such searches cannot probe parameter space corresponding to the canonical QCD axion. Regardless, these interactions represent qualitatively different types of signatures.

\subsection{Fermion Spin}
\label{sec:spin}

%
\begin{figure}[t]
\includegraphics[width=0.85\textwidth]{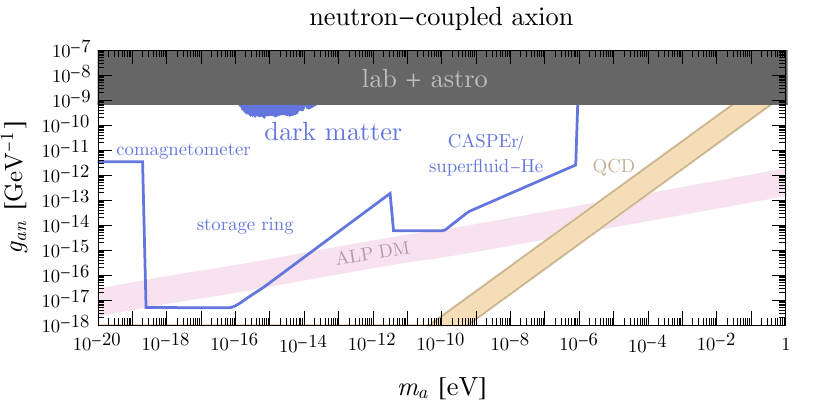}
\caption{As in \Fig{axionphoton}, but for neutron-coupled axions (\Eq{Lagaff} with $f = n$), in the parameter space spanned by the axion-neutron coupling $g_{an}$ and the axion mass $m_a$. The orange band denotes the range of couplings for the DFSZ QCD axion. For hadronic QCD axions, such as in the KSVZ model, recent estimates are consistent with $g_{an} = 0$~\cite{GrillidiCortona:2015jxo}.
\label{fig:neutron}}
\end{figure}

Axion dark matter may also couple to Standard Model fermions $f = n, e, \ldots$, as in the last term of \Eq{axioncouplings}, which we rewrite here as\footnote{Although not considered in this review, it is also possible that the axion couples \emph{off-diagonally} to the various fermion generations~\cite{Wilczek:1982rv,Bjorkeroth:2018dzu,MartinCamalich:2020dfe}.}
\be
\label{eq:Lagaff}
\mathcal{L} \supset g_{a f} \, (\partial_\mu a) \, \bar{\psi} \, \g^\mu \g^5 \, \psi
~,
\ee
where $g_{a f} \propto 1 / f_a$. At low-energies, \Eq{Lagaff} gives rise to the following nonrelativistic single-particle Hamiltonian for $f$~\cite{Berlin:2023ubt},
\be
\label{eq:Hamaff}
H \supset - g_{af} \, (\grad a) \cdot \sigmav - (g_{a f} / m_f) \, (\dt a) \, \sigmav \cdot \piv
~,
\ee
where $\piv = \pv - q_f \, \A$ is the mechanical momentum of the fermion with charge $q_f$, $\pv = - i \grad$ is the canonical momentum, and $\A$ the electromagnetic vector potential. The first and second terms of \Eq{Hamaff} are referred to as the ``axion wind" and ``axioelectric" terms. Their effect on Standard Model fermions can be deduced by working out the modifications to the fermion's equation of motion via Ehrenfest's theorem. Note that for the axion wind term, $\grad a$ couples to the fermion's spin analogous to a magnetic field, whereas for the axioelectric term, $(\dt a) \, \sigmav$ couples to the fermion's momentum analogous to a vector potential. 

\begin{figure}[t]
\includegraphics[width=0.85\textwidth]{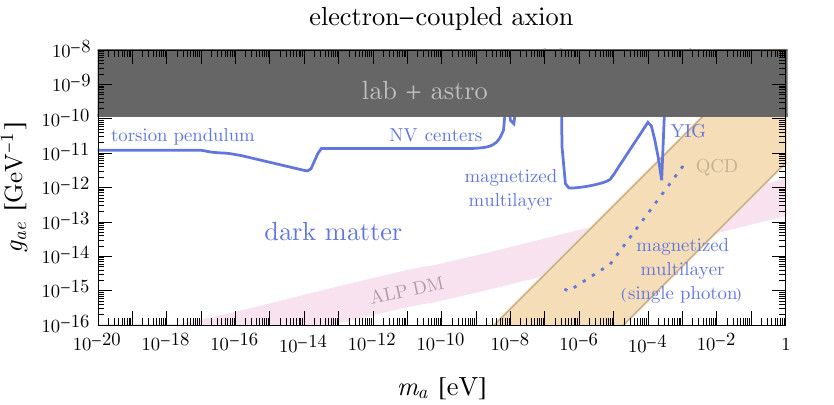}
\caption{As in \Fig{axionphoton}, but for electron-coupled axions (\Eq{Lagaff} with $f = e$), in the parameter space spanned by the axion-electron coupling $g_{ae}$ and the axion mass $m_a$. The orange band denotes the range of couplings for the canonical QCD axion. The dotted blue line shows the maximum possible reach of a magnetized multilayer setup given noise-free detection of single photons~\cite{Berlin:2023ubt}.
\label{fig:electron}}
\end{figure}

As a result, the dominant effect of fermion-coupled axion dark matter is to act with a torque or force on fermion spins, which can be phrased in terms of an effective axion-wind magnetic field and axioelectric electric field, respectively,
\be
\B_\text{eff} = \frac{g_{a f}}{\mu_f} \, \grad a
~~,~~
\E_\text{eff} = - \frac{g_{a f}}{q_f} \, \frac{d}{d t} \, \big( \dot{a} \, \langle \sigmav \rangle\big)
~,
\ee
where $\mu_f$ is the Bohr or nuclear magneton for leptons or nucleons~\cite{Berlin:2023ubt}. For typical experimental setups in which $d \langle \sigmav \rangle  / d t = 0$, $\E_\text{eff} \propto \ddot{a}$ decouples more rapidly at small axion masses, such that $\B_\text{eff}$  leads to larger signals at low $m_a$. Existing limits and projected sensitivities are shown for the axion-neutron coupling and axion-electron coupling in Figs.~\ref{fig:neutron} and \ref{fig:electron}.

Let us first focus on experimental schemes to detect the axion wind term. The effective magnetic field $B_\text{eff}$ causes spins (nucleons or electrons, depending on the fermion coupled to the axion) to precess about $\grad a$, which is set by the direction of the dark matter wind $\vv_\DM$. Nuclear spins are best suited for low-frequency interactions, since they tend to be well-shielded from external degrees of freedom and their corresponding Larmor frequency is suppressed by the nucleon mass. Indeed,  at sub-$\mu \eV$ frequencies, many precision magnetometry techniques may be immediately repurposed as an axion search, including spin-polarized torsion pendulums~\cite{Terrano:2019clh,Graham:2017ivz}, nuclear magnetic resonance experiments~\cite{Graham:2013gfa,Graham:2017ivz,Abel:2017rtm,Xu:2023vfn}, comagnetometers~\cite{Wu:2019exd,Bloch:2021vnn,Bloch:2022kjm,Lee:2022vvb}, and hybrid spin resonance systems~\cite{Wei:2023rzs}. Recently, there have also been proposals involving alternative techniques such as spin precession in storage rings~\cite{Graham:2020kai,Janish:2020knz,JEDI:2022hxa}, superfluid $^{3}$He in both the A phase~\cite{Chigusa:2023szl} and B phase~\cite{Gao:2022nuq,Foster:2023bxl}, and nitrogen vacancy centers~\cite{Chigusa:2023hms}. 

For $m_a \gtrsim 1 \ \mu \eV$, the axion wind can couple to electron spin-excitations with larger characteristic energy. Such examples include spin-flip transitions in atoms~\cite{Sikivie:2014lha} and in-medium magnon excitations~\cite{Mitridate:2020kly}. Since collective spin-excitations source electromagnetic fields, the most well-defined readout scheme for magnons are electromagnetic ones. For instance, certain experiments employ electromagnetic readout by placing an electron spin-polarized sample in a cavity, mixing the magnon and cavity modes~\cite{Barbieri:2016vwg,QUAX:2020adt,Ikeda:2021mlv}. In order to leverage the high-$Q$ of RF cavities, these setups employ low-loss magnetic insulators, such as yttrium iron garnet (YIG). However, such detectors are typically limited in exposure due to the fact that YIG crystals are difficult to manufacture beyond the mm-scale. Alternatively, one could incorporate cheaper magnetic materials, such as  polycrystalline spinel ferrites, which can be scaled up to a much larger volume at the cost of increased loss. In this case, dielectric layers analogous to the MADMAX experiment discussed in \Sec{HF}, can be utilized, but to be sensitive to the axion-electron coupling, the dielectric needs to consist of magnetic material. Such a setup, referred to as a ``magnetized multilayer," was proposed in Ref.~\cite{Berlin:2023ubt}, involving sensitive photon detectors to probe unexplored parameter space at $\gtrsim \GHz$ frequencies. More generally, for most experimental setups targeting the axion wind term, the total axion absorption power is governed by the \emph{magnetic} energy-loss function, $P_\text{sig} \propto \vv_\DM \cdot \text{Im} (- \hat{\boldsymbol{\mu}}^{-1}) \cdot \vv_\DM$, where $\hat{\boldsymbol{\mu}}$ is the contribution of the spin of the fermion $f$ to the medium's permeability tensor (i.e., for electrons this is the usual magnetic permeability) evaluated at the angular frequency $m_a$~\cite{Berlin:2023ubt}. Thus, the dissipative part of the target permeability completely determines the inclusive dark matter absorption rate.

Compared to the axion wind, the possible observables arising from the axioelectric term have not been as closely examined. Most effects can be understood from the corresponding effective electric field $\E_\text{eff}$, which acts as a force aligned along the spin of the fermion~\cite{Berlin:2023ubt}. For instance, for electron-coupled axions, the axioelectric term can induce modifications to atomic energy levels, although such effects are typically higher order in the electromagnetic fine-structure constant~\cite{Berlin:2023ubt}. $\E_\text{eff}$ can also acts as a bulk mechanical force in spin-polarized material, drive electronic excitations in molecules~\cite{Arvanitaki:2017nhi}, superconductors~\cite{Hochberg:2016ajh}, semiconductors~\cite{Hochberg:2016sqx}, and materials hosting optical phonons~\cite{Mitridate:2021ctr}, as well as induce polarization currents in dielectric stack experiments employing electron-polarized materials~\cite{Berlin:2023ubt}. 

\section{Producing and Detecting New Particles in the Lab}
In Sec.~\ref{sec:DM}, we focused on experiments searching for galactic dark matter present in the laboratory. Using the same interactions enumerated in Sec.~\ref{sec:theory}, we may also produce the axion or dark photon fields in the laboratory, and subsequently reconvert them to detectable laboratory fields or forces. Such ``light-shining-through-wall" experiments are generally more difficult than dark matter experiments because production and detection requires additional powers of small couplings. On the other hand, light-shining-through-wall experiments do not rely on dark matter being present in the lab, and are thus a powerful probe of \emph{any} axion or dark photon in the Lagrangian of the universe, independent of any astrophysical uncertainties. As with dark matter experiments, the most straightforward detection strategies are electromagnetic, with new technologies recently providing vastly improved sensitivities. Axions may also lead to spin-dependent forces which can be detected with precision magnetometry and mechanical sensing.

\subsection{Electromagnetic}

Light-shining-through-wall experiments aim to both create and detect new particles, such as electromagnetically-coupled axions or dark photons. After first being sourced by the ``emitter" (a region of strong electric $E_\text{em}$ and/or magnetic $B_\text{em}$ fields), these particles then propagate into a quiet shielded ``receiver" region, in which they can excite small signal electromagnetic fields. The degree to which axions or dark photons are sourced by the emitter can be determined from their equations of motion, which are, respectively
\begin{align}
& (\partial^2 + m_a^2) \, a  = \gagg \, \E \cdot \B
\label{eq:axionEOM}
\\
& \bigg( \partial^2 + \frac{\mAp^2}{1-\eps^2} \bigg) \, A_\text{inv}^\mu = - \frac{\eps \,  \mAp^2}{1 - \eps^2} \, \, A^\mu
\label{eq:ApEOM}
~,
\end{align}
where on the left-hand side of \Eq{ApEOM} we have worked in the basis introduced above \Eq{jeffAprime}. 
From this, we see that axions are sourced by regions in which an electric \emph{and} magnetic field are aligned, whereas dark photon fields can be produced from either electric or magnetic fields. Regardless, the ideal emitter involves fields driven to high intensity, which can be achieved through the use of, e.g., optical or RF cavities. Furthermore, from the form of the effective currents in \Eqs{jeffaxion}{jeffAprime}, detecting the emitted axions requires an additional magnetic field $B_\text{rec}$ in the receiver, whereas detection of emitted dark photons does not. 

The figure of merit for these experiments is the probability for photons to convert into an axion or dark photon and then reconvert back into photons in the receiver. In optimal cavity setups operating at frequencies $\w_\text{exp}$ much greater than the axion or dark photon mass, the conversion probability is typically of the form
\be
\label{eq:conversion}
P_{\g_\text{em} \to \g_\text{rec}} \sim 
\begin{cases}
\big( \gagg \, \sqrt{B_\text{em} \, B_\text{rec}} \, L_\text{exp} \big)^4
& \text{(axion)}
\\
(\eps \, \mAp / \w_\text{exp})^4
& \text{(dark photon)}
~,
\end{cases}
\ee
where $L_\text{exp}$ is an experimentally-dependent length scale, dictated by the size of the emitter and receiver and the distance separating them.In both cases, finesse and/or quality factors can further boost the power in the receiver cavity. A key difference between axion and dark photon experiments is the existence of the longitudinal mode of the dark photon, which can have a parametrically large effect on the signal, depending on the geometry of the experiment. As highlighted in Ref.~\cite{Graham:2014sha}, to obtain the $\mAp^4$ scaling in \Eq{conversion}, the emitter and receiver cavities must be aligned with the direction of $\E_\text{em}$; instead, if the receiver cavity is off-axis, then $P_\text{conv} \propto \mAp^8$, which severely suppresses the signal rate for low-mass ($\mAp \ll \w_\text{exp}$) dark photons.

The strongest laboratory-based limits on dark photons come from the Dark SRF experiment~\cite{Romanenko:2023irv}, which utilizes 1.3 GHz emitter and receiver SRF cavities arranged coaxially to enhance sensitivity to the dark photon's longitudinal polarization. Both cavities have large intrinsic quality factors, $Q_\text{int} \gtrsim 10^{10}$, a significant improvement over previous microwave cavity experiments with $Q \lesssim 10^4$~\cite{Betz:2013dza,Parker:2013fxa}. A first pathfinder run using $E_\text{em} = 6.2 \ \MV / \text{m}$ and a total data-taking time of a few hours observed no excess power in the receiver cavity above thermal noise (following rejection of a peak also observed when the emitter cavity was off), and set a limit of $\eps \lesssim 1.6 \times 10^{-9} \times (5 \ \mu \eV / \mAp)$ for $\mAp \lesssim 5  \ \mu \eV$. This was the first demonstration of high-$Q$ SRF cavities for dark photon detection, including the required temperature and frequency stability. However, the sensitivity of this run was limited by an unwanted frequency offset between the emitter and receiver cavities. Future longer runs of the experiment with better frequency matching and larger $E_\text{em}$ are expected to significantly improve the sensitivity. 

For larger masses, i.e., when $m_a , \mAp \gg \w_\text{exp}$, axions and dark photons can only be sourced off-shell, corresponding to the production of evanescent fields that fall off exponentially within a Compton wavelength from the emitter. For light-shining-through-wall experiments whose emitter and receiver regions are separated by more than $\sim \w_\text{exp}^{-1}$, such signals are exponentially suppressed. This is the case for, e.g., Dark SRF, which has limited sensitivity to masses above $\text{GHz} \sim \mu \eV$. To get around this, Ref.~\cite{Bogorad:2019pbu} proposed using the \emph{same} cavity as both source and receiver by simultaneously pumping two modes, which could offer better sensitivity at large axion masses, but likely suffers from a large irreducible background of nonlinear harmonic generation in the superconducting walls~\cite{Ueki:2024sfk} (which may nonetheless be of some intrinsic interest to condensed matter physicists). However, for masses $\lesssim 1 \ \eV$, the Compton wavelength is still much larger than the electromagnetic penetration depth in superconducting material $\sim 50 \ \text{nm}$. As a result, the emitter and receiver cavities can be separated by much less than a Compton wavelength while also maintaining efficient shielding. Such light-shining-through-\emph{thin}-wall (LSthinW) experiments were recently proposed in Ref.~\cite{Berlin:2023mti}, which showed that setups analogous to Dark SRF could achieve sensitivity to unexplored dark photon parameter space for masses as large as $\mAp \sim \text{few} \times 10 \ \text{meV}$.

The strongest purely laboratory-based axion limits are expected to come from the ALPS II experiment~\cite{Wei:2024fkf}, which consists of 1064 nm light propagating through a string of twelve $5.3 \ \text{T}$ magnets in the source region, followed by a similar configuration in the receiver region, for a total of $B_0 \, L_B = 560 \ {\rm T} \ {\rm m}$ on each side. Mode-matched optical cavities are used to boost both source and receiver power, yielding an expected sensitivity to a conversion probability of $P_{\g_\text{em} \to \g_\text{rec}} \sim 10^{-25}$, i.e., a few photons per day with 40 W of source power. Two complementary detection strategies are envisioned for such weak signals: heterodyne detection by mixing the signal field with a strong local oscillator and performing single-photon counting at the beat frequency, followed by direct measurement of the signal photons with transition-edge sensors at an expected dark rate of $6.9 \times 10^{-6} \ {\rm Hz}$. The first science-run data was collected in 2023, setting limits of $\gagg \lesssim 6 \times 10^{-10} \ \GeV^{-1}$, and design sensitivity is expected to be achieved by 2025, with an expected sensitivity to $\gagg$ of $2 \times 10^{-11} \ \GeV^{-1}$. One may also envision using SRF cavities as source and receiver cavities~\cite{Bogorad:2019pbu,Janish:2019dpr,Gao:2020anb,Salnikov:2020urr}, with similar projected sensitivities as ALPS II but with different noise considerations.

Finally, we briefly mention helioscope experiments, which aim to detect axions produced from the Sun by converting them to keV X-ray photons with a spectrum determined by the solar temperature. The setup works similarly to ALPS II but without the source cavity, and thus the conversion probability is $(\gagg \, B_\text{rec} \, L_\text{exp})^2$ for $m_a \ll \text{keV}$. CAST~\cite{CAST:2017uph}, which consists of a $9 \ \text{T}$ LHC dipole magnet on a rotating mount to follow the direction of the Sun, completed data-taking in 2015 and set the strongest laboratory limits of $\gagg < 6.6 \times 10^{-11} \ \GeV^{-1}$ over many orders of magnitude in axion mass, which have persisted for almost a decade. The IAXO pathfinder~\cite{CAST:2024eil} has implemented a new conversion gas to slightly improve sensitivity in the high-mass region, improving on a similar strategy used by CAST; by tuning the gas pressure, the plasma frequency can be tuned to match the axion mass, which makes the axion-to-photon conversion coherent over the length $L_\text{exp}$ (analogous to the plasma haloscopes discussed in Sec.~\ref{sec:HF}).

\subsection{CP-violating couplings and axion-mediated forces}

Recently, there has been interest in pursuing more exotic couplings, such as when axions possess interactions with Standard Model operators that are instead \emph{even} under parity and time-reversal. As mentioned at the end of \Sec{QCD}, this typically arises at higher order in the axion field, i.e., $\sim \order{a^2/f_a^2}$. This is indeed the case for minimal models of the QCD axion. To see this, note that the total effective QCD theta angle is $\theta_\text{tot} = \bar{\theta} + \langle a \rangle / f_a + a / f_a$, where $\bar{\theta}$ is the bare value, $\langle a \rangle$ is the axion vacuum expectation value, and $a$ is the fluctuation around this value (e.g., in the case of dark matter $a / f_a = \theta_a$ in \Eq{thetaQCD}). When the only source of parity and time-reversal violation is $\bar{\theta} \neq 0$, the energetic minimum is achieved at $\langle a \rangle / f_a = - \bar{\theta}$, such that $\theta_\text{tot} = a / f_a$~\cite{Vafa:1984xg}. Since parity and time-reversal symmetric operators couple only to the square of $\theta_\text{tot}$ at leading order, the same is true for $a / f_a$.

However, if there are additional sources of parity and time-reversal violation beyond  $\bar{\theta}$, these can induce $\langle a \rangle / f_a \neq - \bar{\theta}$, such that $\theta_\text{tot} = \theta_\text{eff} + a / f_a$, where $\theta_\text{eff} \equiv \bar{\theta} + \langle a \rangle / f_a \neq 0$. While the Standard Model contribution to $\theta_\text{eff}$ is experimentally negligible, additional new physics may lead to values that saturate the current experimental limit of $\theta_\text{eff} \lesssim 10^{-10}$. Scalar interactions controlled by $\theta_\text{tot}^2$ therefore involve the cross product $\theta_\text{eff} \times (a / f_a)$. For example, such considerations imply the existence of scalar couplings between the axion and Standard Model fermions parametrically of the form $\mathcal{L} \sim \theta_\text{eff} \, (m_f/f_a) \, a \, \bar{\psi} \psi$, where $m_f$ is the mass of the fermion~\cite{Moody:1984ba} (see also, e.g., Refs.~\cite{Bertolini:2020hjc,DiLuzio:2024ctr} for recent discussions of these points). In this case, the axion can be sourced by matter density alone, and then may be detected with, e.g., its standard coupling to electromagnetism or spin. Various schemes along these lines have been proposed using spin probes for axions produced by laboratory sources~\cite{Arvanitaki:2014dfa} or the Earth~\cite{Agrawal:2022wjm,Agrawal:2023lmw}.

In the case of laboratory sources, one physical effect is that of a spin-dependent axion-mediated force~\cite{Arvanitaki:2014dfa}, where a source mass creates an effective magnetic field which can in turn be detected with NMR techniques, like those reviewed in Sec.~\ref{sec:spin}. The ARIADNE collaboration~\cite{ARIADNE:2017tdd,ARIADNE:2020wwm} aims to use tungsten as the source mass and polarized $^{3}{\rm He}$ as the detection medium, with projected sensitivity to the QCD axion in the $(10^{-3} - 10^{-5})  \ \eV$ mass range. 

\section{Looking Forward}

As we have surveyed in this review article, the creativity and interdisciplinarity exhibited by the axion and dark photon community is enormous, with dozens of new experiments proposed in the decade since the last such review~\cite{Graham:2015ouw}. Without discounting the enormous effort represented by this cumulative body of work, no discovery of axion or dark photon dark matter has yet been made. In that light, it is worth evaluating the progress of experiments over the past decade in an effort to identify which regimes of parameter space still remain unprobed, which technological aspects have been bottlenecks to progress, and where effort might best be spent to fully probe the best-motivated dark matter models.

The most rapid progress has been made in experiments searching for the axion electromagnetic coupling. While several new searches have obtained sensitivity in narrow frequency ranges, often dipping down into the canonical QCD axion parameter space, a broad program covering the QCD axion over several decades of mass remains an important goal for the future. For cavity searches, the principal technological challenge is the $B_0^4 \, V_\text{exp}^2$ scaling of the scan rate in \Eq{scanrate}. For a single-mode haloscope, achieving DFSZ sensitivity over the $(1-10) \ \GHz$ range with current $B$-field strengths and $Q_L \sim 10^6$ would take 20,000 years of scanning at 100\% livetime~\cite{PalkenThesis}. The technology we described in this review -- including squeezing and photon counting to push beyond the standard quantum limit, and SRF cavities with extremely large $Q_L$, both of which have been experimentally demonstrated in the past decade -- may reduce this daunting task to a more manageable level. However, the most rapid improvements may come with development of high-field magnets, which is strongly synergistic with the fusion energy program~\cite{ChaudhuriPXS}. 

We further emphasize that the $\rhodm^2$ scaling of the scan rate from \Eq{scanrate} (given our uncertainty on the dark matter density and substructure) implies that a comprehensive search strategy must include probing couplings \emph{below} those corresponding to the ``standard'' value of $\rhodm \simeq 0.45 \ \GeV/\cm^3$, a value which is primarily adopted for convenience in comparing experiments rather than a quantity of intrinsic physical significance.

At sub-GHz frequencies, initial development is underway to manufacture tunable lumped-element circuits and high-$Q$ SRF cavities with nearly degenerate modes. However, the importance of noise sources that are more prevalent in this regime, such as low-frequency vibrational noise~\cite{Ouellet:2019tlz}, remains to be seen. If such noise sources can be mitigated, these approaches provide a plausible path to QCD axion detection for masses corresponding to GUT-scale decay constants. At high frequencies, SNSPDs have demonstrated excellent performance in the optical and near-infrared frequency range, but further work is needed to demonstrate a viable path toward QCD axion detection in the ``THz gap,'' which may benefit from the synergy with development of single-photon THz sensors for astrophysics applications~\cite{THzKID,PhysRevX.14.041005}.

By contrast, the parameter space for spin-coupled axion dark matter looks much the same as it did a decade ago. The only experiments whose exclusion limits dip beyond the long-standing neutron star cooling constraints operate in a broadband mode at low frequencies, one of which actually used archival data first taken in 2009 for other purposes~\cite{Lee:2022vvb}. The situation for the defining gluon coupling of the QCD axion is perhaps even starker, with broadband experiments only able to probe axions below 100 Hz, and resonant experiments limited to narrow ``fingers'' of parameter space which do not yet improve upon SN1987A constraints. As outlined in \Sec{QCD}, the gluon coupling can be probed with electromagnetic sensors coupled to nuclear spin-polarized material; however, such strategies are typically hindered by the fact that the QCD axion generates an EDM current much smaller than the effective current from its photon coupling, $J_\text{EDM} / J_{a \g} \lesssim 10^{-3} \times (10 \ \text{T} / B_0)$~\cite{Berlin:2022mia}. New and creative experimental approaches are thus needed to fully explore these well-motivated regions of parameter space.

The search for dark photons will benefit from the continued development of axion experiments, as many of the projections shown in \Fig{darkphoton} correspond to axion experiments without the direct use of the magnetic field. While there is no sharp target in mass-coupling space analogous to the QCD axion, we encourage the experimental community to continue pursuing these searches. In particular, we encourage axion experiments to publish dedicated dark photon searches. With regards to light-shining-through-wall experiments, Dark SRF clearly demonstrates the benefits of high-$Q$ SRF cavities for dark photon detection, and ALPS II should soon achieve its design sensitivity for axions.

To close, we commend once again the combined creativity and efforts of theorists and experimentalists to vastly expand the suite of experiments searching for the axion and dark photon. New technologies have been brought to bear on this problem over the past decade, many of which have shown the potential for technology transfer to physics from fields as diverse as microwave engineering, precision magnetometry, and quantum sensing. While a discovery of new physics is never assured, the field has always progressed by looking in new places, and we look forward to the new and insightful approaches to be proposed and implemented in the decade to come.

\section*{DISCLOSURE STATEMENT}
The authors are not aware of any affiliations, memberships, funding, or financial holdings that
might be perceived as affecting the objectivity of this review. 

\section*{ACKNOWLEDGMENTS}

We would like to sincerely thank Joshua Foster and Noah Kurinsky for helpful discussions. We are particularly grateful to Alex Millar and Kevin Zhou for providing comments on the manuscript, and to Ciaran O'Hare for the \texttt{AxionLimits} repository of axion experimental results and projections~\cite{AxionLimits}. We are indebted to Gianpaolo Carosi for invaluable comments and detailed discussions during the preparation of this review. Fermilab is operated by the Fermi Research Alliance, LLC under Contract DE-AC02-07CH11359 with the U.S. Department of Energy. This material is based upon work supported by the U.S. Department of Energy, Office of Science, National Quantum Information Science Research Centers, Superconducting Quantum Materials and Systems Center (SQMS) under contract number DE-AC02-07CH11359. YK is supported in part by DOE grant DE-SC0015655. This work was completed in part at the Perimeter Institute. Research at Perimeter Institute is supported in part by the Government of Canada through the Department of Innovation, Science and Economic Development Canada and by the Province of Ontario through the Ministry of Colleges and Universities. 

%
\bibliography{bibliography}
 
\end{document}